\documentclass[twocolumn,aps,nofootinbib]{revtex4}

\usepackage{graphicx}
\usepackage{epstopdf}
\usepackage{latexsym}
\usepackage{amsmath}
\usepackage{amssymb}
\usepackage{color}
\usepackage{mathrsfs}

\usepackage[center]{subfigure}

\begin{document}

 \newcommand{\bq}{\begin{equation}}
 \newcommand{\eq}{\end{equation}}
 \newcommand{\bqn}{\begin{eqnarray}}
 \newcommand{\eqn}{\end{eqnarray}}
 \newcommand{\nb}{\nonumber}
 \newcommand{\lb}{\label}
 \newcommand{\be}{\begin{equation}}
\newcommand{\en}{\end{equation}}
\newcommand{\PRL}{Phys. Rev. Lett.}
\newcommand{\PL}{Phys. Lett.}
\newcommand{\PR}{Phys. Rev.}
\newcommand{\CQG}{Class. Quantum Grav.}

\title{Echoes of axial gravitational perturbations in stars of uniform density}

\author{Kai Lin $^{a,b}$}\email{lk314159@hotmail.com}
\author{Wei-Liang Qian $^{b,c,d}$}\email{wlqian@usp.br}

\affiliation {a) Hubei Subsurface Multi-scale Imaging Key Laboratory, Institute of Geophysics and Geomatics, China University of Geosciences, Wuhan 430074, Hubei, China}
\affiliation {b) Escola de Engenharia de Lorena, Universidade de S\~ao Paulo, 12602-810, Lorena, SP, Brazil}
\affiliation {c) Faculdade de Engenharia de Guaratinguet\'a, Universidade Estadual Paulista, 12516-410, Guaratinguet\'a, SP, Brazil}
\affiliation {d) Center for Gravitation and Cosmology, College of Physical Science and Technology, Yangzhou University, Yangzhou 225009, China}

\date{\today}

\begin{abstract}
This work investigates the echoes in axial gravitational perturbations in compact objects.
To this end, we propose an alternative scheme of the finite difference method implemented in two coordinate systems, where the initial conditions are placed on the axis of the tortoise coordinate with appropriate boundary conditions that fully respect the causality.
The scheme is then employed to study the temporal profiles of the quasinormal oscillations in the Schwarzschild black hole and the uniform-density stars.
When presented in a two-dimensional evolution profile, the resulting ringdown waveforms in the black hole metric are split into the reflected and transmitted waves as the initial perturbations evolve and collide with the peak of the effective potential.
On the other hand, for compact stars, quasinormal oscillations might be characterized by echoes.
Consistent with the causality arguments, the phenomenon is produced by the gravitational waves bouncing between the divergent potential at the star's center and the peak of the effective potential.
The implications of the present study are also discussed.
\end{abstract}


\maketitle

\section{Introduction}
Compact objects, such as a neutron star or black hole, give rise to gravitational waves (GWs) triggered by the metric perturbations.
The dissipative process is essentially featured by three stages: the initial burst, the quasinormal mode (QNM) oscillations, and the late-time tail.
In the initial burst stage, the system's dynamical properties are determined mainly by the initial conditions.
On the other hand, in the QNM and late-time tail stages, the process is primarily governed by the intrinsic nature of compact objects.
For spherically symmetric metrics, the master equation can be simplified, and in the frequency domain, the radial sector can be expressed as a Sch\"ordinger type equation regarding a complex eigenfrequency $\omega$.
A variety of methods were proposed, and one is expected to extract valuable information on the underlying stars or black holes from the GW measurements~\cite{LIGO1, LIGO2, LIGO3, LIGO4, LIGO5, LIGO6, LIGO7, LIGO8, LIGO9}.
The relevant studies performed through the analysis of observed quasinormal frequencies, known as QNM spectroscopy, have gained much momentum recently~\cite{QNMmore1, QNMmore2, QNMmore3, QNMmore4, QNMmore5, QNMmore6, QNMmore7}.

Regarding the black hole QNMs, the dynamic temporal evolution of the initial perturbations is usually explored at a given spatial point.
However, it is more meaningful to study the GW propagation for the entire range of spatial coordinates, for instance, in the two-dimensional $t-r_*$ coordinate space (where $r_*=\int dr/\sqrt{fh}$ is the tortoise coordinate).
In the literature, many semi-analytic methods, such as the WKB approximation~\cite{WKB1,WKB2,WKB3,WKB4,WKB5,WKB6,Konoplya1}, the continued fractional method~\cite{CFM1,CFM2}, the asymptotic iteration method\cite{AIM1,AIM2,AIM3,AIM4}, the matrix method \cite{Lin1,Lin2,Lin3,Lin4,Lin5}, among others~\cite{HH,Expand1,NewM1,NewM2,NewM3,NewM4}, have been proposed.
These approaches have been extensively used for evaluating the black hole QNM frequencies $\omega$.
For the star QNMs, however, they must be modified accordingly and often used in conjunction with the shooting method~\cite{Star1, Star2, Star3, Star4, Star5, Star6, Star7}.
Nonetheless, the above approaches only concern the QNM oscillation stage, and therefore they cannot be employed to address the entire process.
In this regard, the finite difference method (FDM) is a suitable approach that can be used to provide a more general description of the dynamical evolution for given initial perturbations~\cite{FDM1, FDM2, FDM3, FDM4, FDM5, FDM6}.

In practice, to implement the FDM, the boundary condition and initial conditions are often proposed on the axes of the Eddington-Finkelstein coordinates, $\Psi(u,v=0)=0$ and $\Psi(u=0,v)=\Psi_0$ (where $u=t-r_*$ and $v=t+r_*$), respectively.
By assuming that the spacetime in the tortoise coordinate is mainly flat, so that the speed of light of the waveform $c\sim 1$, one concludes that the boundary placed on $u=0$ is mostly in accordance with the causality that will never be attained by the initial perturbations assigned to the surface $v=0$.
In practice, the assumption that $c\sim 1$ has been manifestly shown as a reasonable approximation in numerical calculations.
In principle, however, it might be violated in the vicinity of the black hole.
In other words, although the above boundary condition is asymptotically correct, for a finite period, the initial perturbations might temporarily travel across the $v$ axis.
In this regard, we propose an alternative approach as follows.
We place the initial conditions on the spatial surface $t=0$ and adopt appropriate boundary conditions in accordance with the causality.
For the black hole spacetime, free boundary conditions are implemented so that the initial perturbations will propagate as an ingoing wave towards the horizon and an outgoing wave towards spatial infinity.
For the compact star spacetime, the free boundary condition will be adopted at the spatial infinity for the reason mentioned above, and the perturbations must vanish at the center of the star as the effective potential goes to infinity.
Moreover, the junction condition in terms of a vanishing Wronkian is enforced at the star's surface.
The proposed scheme is more flexible in the sense that it can be easily modified to adopt the case where the speed of light exceeds the unit.
Therefore, it is adopted in the present work to investigate the temporal profiles of both black holes and stars.
The obtained results are illustrated in a two-dimensional evolution profile.
In the case of black holes, it is found that the resulting ringdown waveforms are featured by the reflected and transmitted waves as the initial perturbations evolve and collide with the maximum of the effective potential.
On the other hand, in the case of compact stars, temporal evolutions may also be characterized by echoes.
The echoes have largely been speculated to be associated with compact objects~\cite{Cardoso:2016rao, Mark:2017dnq, Bueno:2017hyj, Liu:2021aqh}, such as black holes and wormholes, which has become an intriguing topic in recent years.
Typically, their emergence is understood to be related to the existence of an effective potential well.
Specifically, for the wormhole metric explored in Ref.~\cite{Bueno:2017hyj}, echoes are intuitively attributed to the potential well between the two maxima.
Besides, the echo's period is shown to be governed by twice the distance between the peaks.
In the context of a black hole or other exotic compact objects, the period of the echoes is related to the distance between the maximum of the effective potential and the surface or inner boundary of the compact object~\cite{Mark:2017dnq}.
More recently, it was pointed out that a discontinuity may also play such a role in triggering echoes~\cite{Liu:2021aqh}.
For the present scenario, the echoes can be attributed to the repeated bouncing of the waveforms between the star's center and its surface.

The remainder of the paper is organized as follows.
In the following section, we present the numerical schemes of the FDM and the associated von Neumann stability condition.
In Sec.~III, the obtained numerical results are discussed.
Sec.~IV is devoted to the concluding remarks.

\section{The FDM schemes}

For the sake of simplicity, we consider the spherically symmetric spacetimes, whose metric possess the form
\bqn
\lb{1}
ds^2=-h(r)dt^2+\frac{dr^2}{f(r)}+r^2\left(d\theta^2+\sin^2\theta d\varphi^2\right) .
\eqn

For the Schwarzschild black hole, we have
\bqn
\lb{2}
f(r)=h(r)=1-\frac{2M_S}{r} ,
\eqn
where $M_S$ and $r_p=2M_S$ are the mass and horizon of the black hole.
The tortoise coordinate $r_*=\int dr/\sqrt{fh}$ reads $r_*=r+r_S\log\left(\frac{r}{r_S}-1\right)$.

For the stars with uniform density $\rho=\frac{3M_S}{4\pi r_b^3}$, where $r_b$ is the radial coordinate of the star's surface, the metric is idential to Eq.~\eqref{2} for the outside region $r \ge r_b$.
On the inside of the star ($r<r_b$), one has~\cite{Weinberg}
\bqn
\lb{3}
f(r)&=&1-\frac{2M(r)}{r}\nb ,\\
h(r)&=&\frac{1}{4}\left(3\sqrt{1-\frac{2M(r)}{r}}-\sqrt{1-\frac{2M(r)r^2}{r_b^3}}\right)^2\nb ,\\
M(r)&=&\int^r_04\pi r'^2 \rho dr'=M_S\frac{r^3}{r_b^3}\nb ,\\
p(r)&=&-\rho\left(\frac{\sqrt{1-\frac{2M(r)}{r}}-\sqrt{1-\frac{2M(r)r^2}{r_b^3}}}{3\sqrt{1-\frac{2M(r)}{r}}-\sqrt{1-\frac{2M(r)r^2}{r_b^3}}}\right) .
\eqn
For the tortoise coordinate, the constant of integration is chosen so that $r_*(r=0)=0$.

By using the method of seperation of variables, the axial gravitational perturbations are governed by~\cite{masterEquation}
\bqn
\lb{4}
&&\frac{\partial^2\Psi}{\partial r^2}-\frac{\partial^2\Psi}{\partial t^2}-V(r)\Psi=0\nb ,\\
&&V_\text{star}(r)=\frac{h}{r^3}\left(L(L+1)r+4\pi(\rho-p)r^3-6M\right)\nb ,\\
&&V_\text{BH}(r)=\frac{h}{r^3}\left(L(L+1)r-6M_S\right) .
\eqn
At spatial infinity, the boundary condition dictates that the waveform $\Psi$ is a symptotic outgoing wave
For the black holes, $\Psi$ must be an asymptotic ingoing wave at the horizon.
For the stars, wave function must be regular at the center of the star and the junction conditions at the star's surface reads
\bqn
\lb{5}
\Psi_\text{inside}(r_b)&=&\Psi_\text{outside}(r_b)\nb ,\\
\partial_r\Psi_\text{inside}(r_b)&=&\partial_r\Psi_\text{outside}(r_b)\nb ,\\
\text{or}~~~~\partial_{r_*}\Psi_\text{inside}(r_b)&=&\partial_{r_*}\Psi_\text{outside}(r_b) ,
\eqn
if the effective potential is not divergent at the surface.

In the proposed scheme, as shown in the left plots of FIG.~1 and 2, the initial conditions are given at the spatial surface $t=0$, namely, $\Psi(t=0)$ and $\partial_t\Psi(t=0)$.
One proceeds to discretize the spacetime coordinates in $r_*$ and $t$ and approximates the partial derivatives by first-order finite differences.
To be specific, we denote $t_i=t_0+i\Delta t$, $r_{*j}=r_{*0}+j\Delta r_*$ and $\Psi^i_j=\Psi(t=t_i,r_*=r_{*j})$, $V_j=V(r_*=r_{*j})$, and the master equation Eq.~\eqref{4} becomes
 \bqn
\lb{6}
\Psi^{i+1}_j&=&-\Psi^{i-1}_j+\frac{\Delta t^2}{\Delta r_*^2}\left(\Psi^i_{j-1}+\Psi^{i-1}_j\right)\nb \\
&&+\left(2-2\frac{\Delta t^2}{\Delta r_*^2}-\Delta t^2V_j\right)\Psi^i_j .
\eqn
For the stars, the junction condition Eq.~\eqref{5} reads
\bqn
\frac{\Psi^i_{j_b}-\Psi^i_{j_b-1}}{\Delta r_*}=\frac{\Psi^i_{j_b+1}-\Psi^i_{j_b}}{\Delta r_*} , \nb
\eqn
where $\Psi^i_{j_b-1}=\Psi_\text{inside}(r_*=r_{*0}+(j_b-1)\Delta r_*)$, $\Psi^i_{j_b+1}=\Psi_\text{outside}(r_*=r_{*0}+(j_b+1)\Delta r_*)$,
and $\Psi^i_{j_b}=\Psi_\text{inside}(r_*=r_{*0}+j_b\Delta r_*)=\Psi_\text{outside}(r_*=r_{*0}+j_b\Delta r_*)$, so that
 \bqn
\lb{7}
\Psi^i_{j_b}=\frac{1}{2}\left(\Psi^i_{j_b+1}+\Psi^i_{j_b-1}\right) ,
\eqn
where the subscript $b$ indicates that the grid is on the star's surface.

The temporal evolution is performed by iterating from both boundaries towards the center.
Usually, Eq.~\eqref{6} is utilized to determine the grid values for the next time step, except for those on the star's surface where the iteration formula would involve grids on both sides of the surface.
For the latter case, one must use Eq.~\eqref{7} instead to find the value of the wave function on the surface, then continue to proceed with Eq.~\eqref{6}.
To avoid the von Neumann instability~\cite{FDM4,FDM5}, we choose $\frac{\Delta t^2}{\Delta r_*^2}+\frac{\Delta t^2}{4}V_\text{max}<1$.

Alternatively, one can also explore the problem in the Eddington-Finkelstein coordinates $u=t-r_*, v=t+r_*$.
The grid distributions are illustrated on the right of FIG.~1 and 2.
Although the discretization of the master equation is carried out in a different coordinate system, as discussed below, it is noted that the essential difference from the conventional approach resides in the boundary conditions.
To be specific, in the case of the black hole metric, the free boundary condition is adopted instead of assigning to the line $v=0$.
In terms of the Eddington-Finkelstein coordinates, the master equation Eq.~\eqref{4} can be rewritten as
\bqn
\lb{8}
\frac{\partial^2\Psi}{\partial u\partial v}+\frac{1}{4}V(r)\Psi=0 .
\eqn
Similarly, one denotes $v_i=v_0+i\Delta v$, $u_j=u_0+i\Delta u$ and $\Psi^i_j=\Psi(v_i,u_j)$, $V^i_j=V(v_i,u_j)$, so the discritized equation is found to be
\bqn
\lb{9}
\Psi^{i+1}_{j+1}=-\Psi^i_j+\left(1-\frac{\Delta^2}{8}V^i_j\right)\left(\Psi^{i}_{j+1}+\Psi^{i+1}_{j}\right) ,
\eqn
where we have considered $\Delta v=\Delta u=\Delta$.
The iteration can be carried out as the value of a grid point is determined by three grid points to the immediate west, south, and south-west of the grid in question.

In the case of a star, again, the above procedure breaks when the iteration involves grids on both sides of the star's surface, which is a straight line of 45 degrees.
By using the relation $\partial_{r_*}=\frac{\partial u}{\partial{r_*}}\partial_u+\frac{\partial v}{\partial{r_*}}\partial_v$, one rewrites the junction condition as
\bqn
\partial_u\Psi_\text{inside}-\partial_v\Psi_\text{inside}=\partial_u\Psi_\text{outside}-\partial_v\Psi_\text{outside} , \nb
\eqn
and therefore its form using finite difference reads
\bqn
\lb{10}
&&\frac{\Psi^{i_b}_{j_b}-\Psi^{i_b}_{j_b-1}}{\Delta}-\frac{\Psi^{i_b+1}_{j_b}-\Psi^{i_b}_{j_b}}{\Delta}\nb\\
&&=\frac{\Psi^{i_b}_{j_b+1}-\Psi^{i_b}_{j_b}}{\Delta}-\frac{\Psi^{i_b}_{j_b}-\Psi^{i_b-1}_{j_b}}{\Delta}\nb ,\\
&&\text{or}~~~\Psi^{i_b}_{j_b+1}+\Psi^{i_b+1}_{j_b}-4\Psi^{i_b}_{j_b}=-\Psi^{i_b}_{j_b-1}-\Psi^{i_b-1}_{j_b} ,
\eqn
where $\Psi^{i_b}_{j_b}=\Psi(v_{i_b},u_{j_b})$ is the grid on the star's surface with radial coordinate $r_b$.
It is readily to confirm that $\Psi^{i_b}_{j_b-1}$, $\Psi^{i_b+1}_{j_b}$ are localed on the inside of the star, and $\Psi^{i_b}_{j_b+1}$, $\Psi^{i_b-1}_{j_b}$ are localed on the outside.
However, the above iteration relation involves unkown grid points $\Psi^{i_b+1}_{j_b}$ and $\Psi^{i_b}_{j_b+1}$.
This can be solved by substituting Eq.~\eqref{9} for those points, namely,
\bqn
\lb{11}
\Psi^{i_b}_{j_b+1}&=&\Psi^{i_b-1}_{j_b}+\Psi^{i_b}_{j_b}-\Psi^{i_b-1}_{j_b}-\frac{\Delta^2}{4}V^{i_b}_{j_b+1}\Psi^{i_b-1}_{j_b}\nb ,\\
\Psi^{i_b+1}_{j_b}&=&\Psi^{i_b}_{j_b}+\Psi^{i_b+1}_{j_b-1}-\Psi^{i_b}_{j_b-1}-\frac{\Delta^2}{4}V^{i_b+1}_{j_b}\Psi^{i_b}_{j_b-1}\nb ,\\
\eqn
into Eq.~\eqref{10} and one finds the desired relation
\bqn
\lb{12}
\Psi^{i_b}_{j_b}&=&\frac{1}{2}\left[\Psi^{i_b-1}_{j_b+1}+\Psi^{i_b+1}_{j_b-1}\right.\nb\\
&&\left.-\frac{\Delta^2}{4}\left(V^{i_b}_{j_b+1}\Psi^{i_b-1}_{j_b}+V^{i_b+1}_{j_b}\Psi^{i_b}_{j_b-1}\right)\right] .
\eqn
We choose $\left|1-\frac{\Delta^2}{16}V_\text{max}\right|<1$ to avoid the von Neumann instability.

\begin{figure*}[tbp]
\centering
\includegraphics[width=0.8\columnwidth]{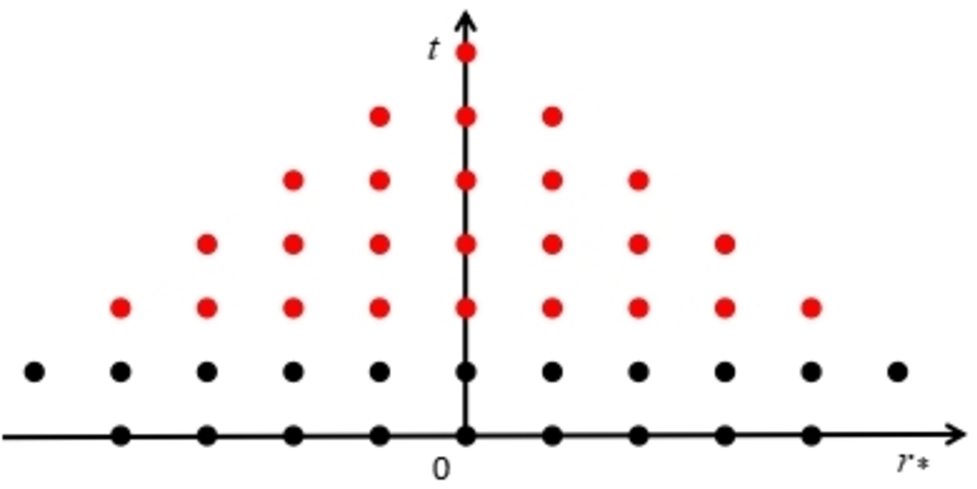}\includegraphics[width=0.8\columnwidth]{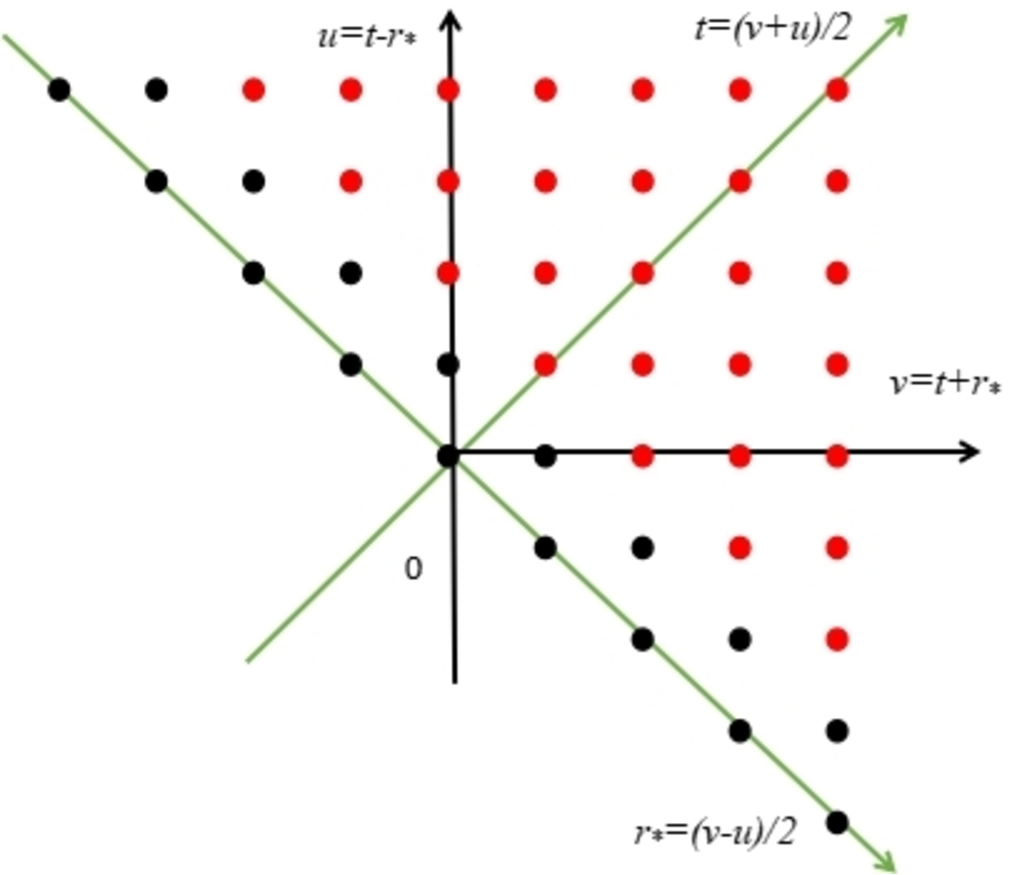}
\caption{The grid layouts of two FDM implementations in the case of black hole.
The left plot corresponds to the proposed scheme in $(r_*, t)$ coordinates, and the right plot is the conventional $(u, v)$ codrdinates.
The black points are the grids to which the initial conditions are assigned, and one uses the iteration process to evaluate the red points, as described in the text.}
\lb{Fig1}
\end{figure*}

\begin{figure*}[tbp]
\centering
\includegraphics[width=0.8\columnwidth]{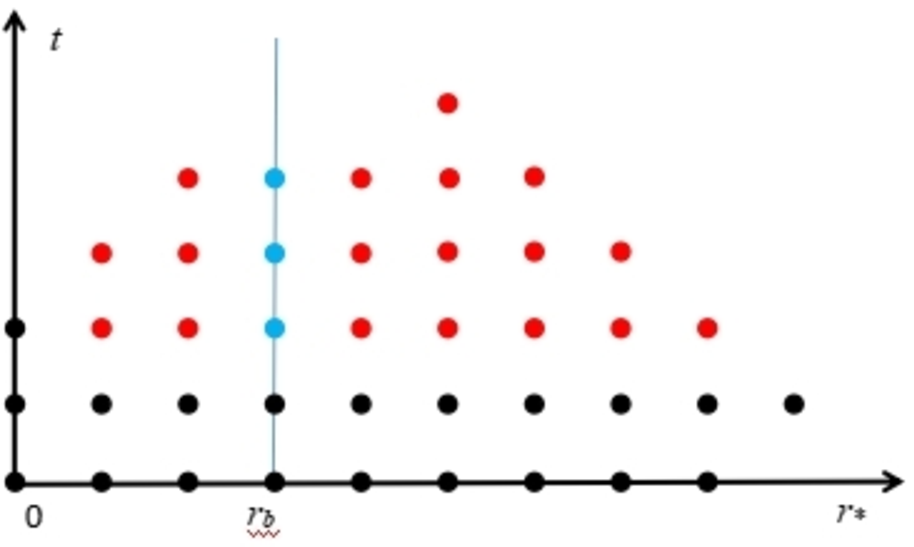}\includegraphics[width=0.8\columnwidth]{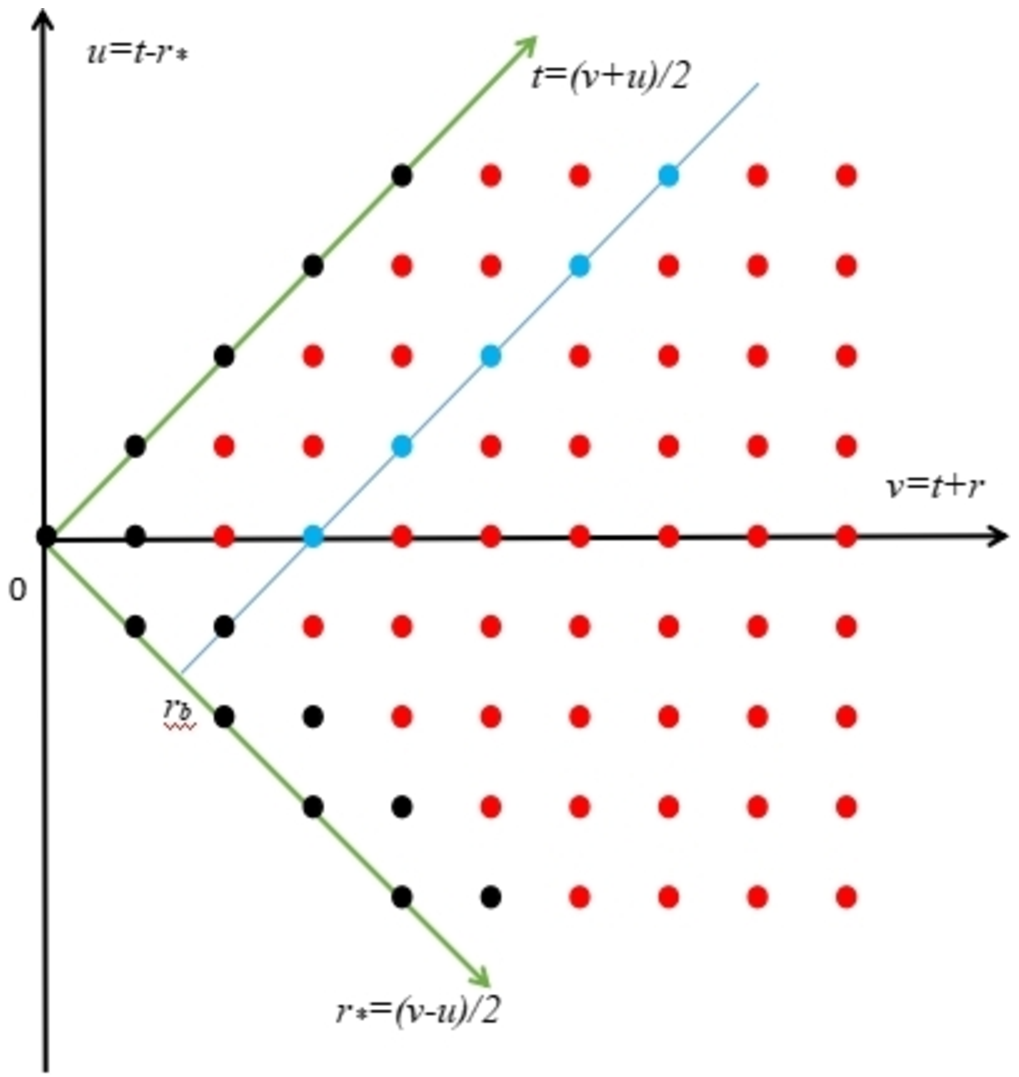}
\caption{The same as FIG.~1 but for the case of star.
Here, the light blue points correspond to the star's surface $r=r_b$, which should be evaluated according to the iteration process is described in the text.}
\lb{Fig2}
\end{figure*}

\section{Numerical results}

Now, we are in a position to present the numerical results using the schemes discussed in the previous section.
The calculations presented below have been carried out for both schemes, and the results are manifestly consistent. 

\begin{figure*}[tbp]
\centering
\includegraphics[width=0.8\columnwidth]{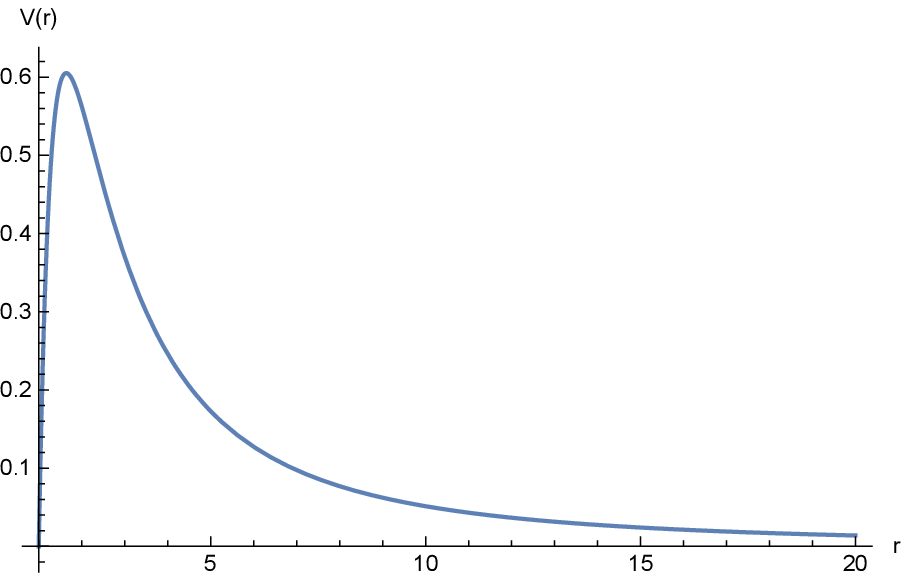}\includegraphics[width=0.8\columnwidth]{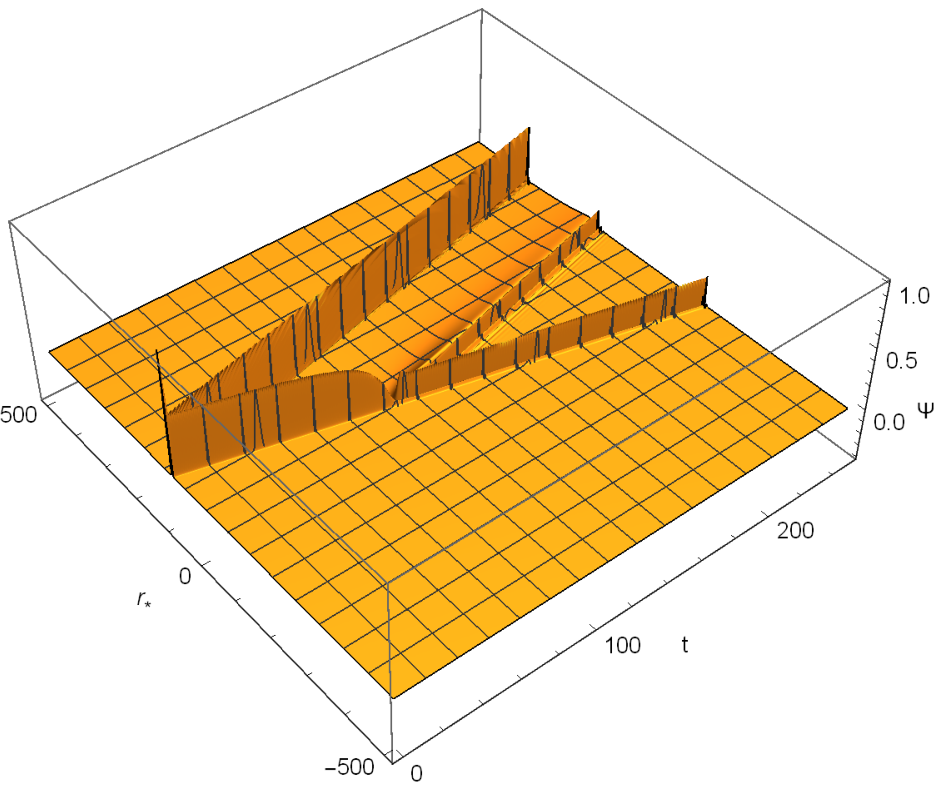}
\includegraphics[width=0.8\columnwidth]{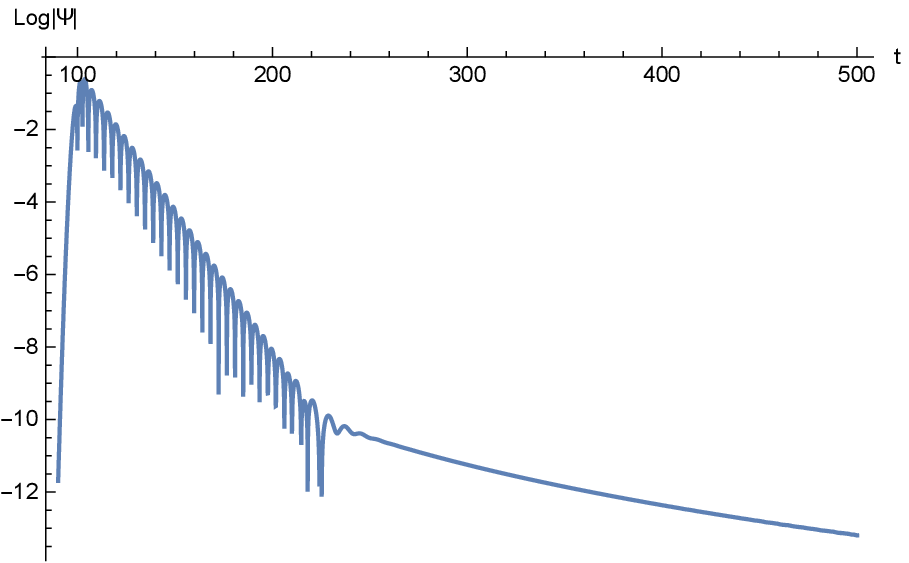}\includegraphics[width=0.8\columnwidth]{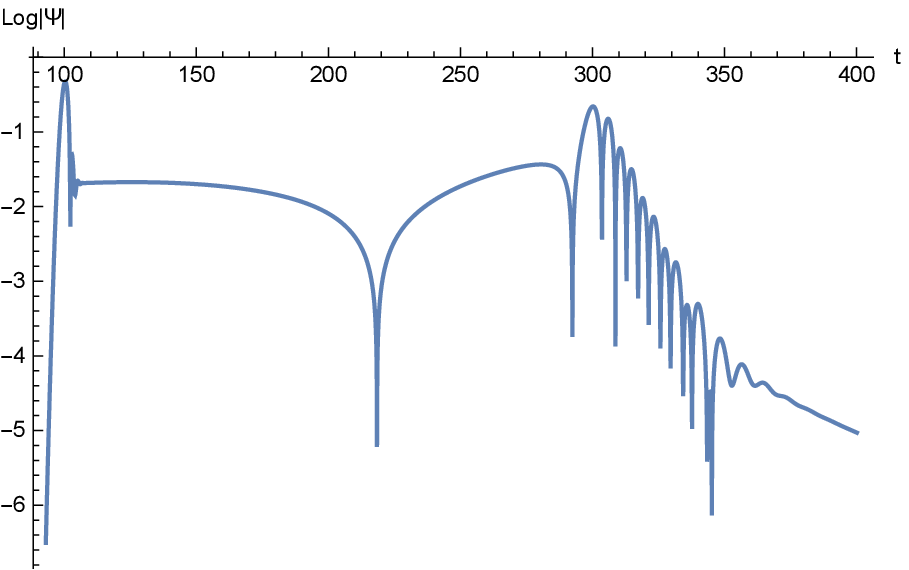}
\caption{The temporal evolution of the axial gravitational perturbations in the Schwarschild black hole with the horizon $r_p=1$.
Top-left: The effective potential. The maximal value of potential is located at $r=1.64039$ ($r_*=1.19471$).
Top-right: The two-dimensional profile of the temporal evolution, where the initial conditions read $\Psi(t=0)=e^{-(r_*-100)^2/2}$ and $\partial_{t}\Psi(t=0)=0$, placed outside of the maximum of the potential.
Bottom-left: The temporal evolution evaluated at $r=1.278$ ($r_*=0$), inside the maximum of the potential.
Bottem-right: The temporal evolution evaluated at $r=194.73$ ($r_*=200$), outside the initial perturbations.}
\lb{Fig3}
\end{figure*}

In FIG.~3, we show the temporal evolutions of the axial gravitational perturbations of the Schwarzschild black hole with the horizon $r_p=1$.
The results are presented in a two-dimensional profile and for given spatial positions.
The initial perturbations are placed on the outside of the potential's peak.
It is observed that the GW propagates in both directions.
The wave that propagates towards the potential's peak is split into reflected and transmitted waves.
The reflected wave propagates to spatial infinity, which eventually gives rise to the quasinormal oscillations, as first pointed out by Andersson~\cite{Andersson}.
On the other hand, the wave that initially propagates outward is associated with the initial burst, namely, the first stage of the dissipative oscillations, as clearly indicated by the bottom-right plot.
It is readily confirmed that the time scales for the occurrence of different stages of the QNM oscillations are in good agreement with the causality arguments.
Moreover, the amplitudes of the GWs numerically satisfy the flow conservation in the asymptotic regions, namely, $|\mathcal{R}^2|+|\mathcal{T}^2|=1$.

\begin{figure*}[tbp]
\centering
\includegraphics[width=0.8\columnwidth]{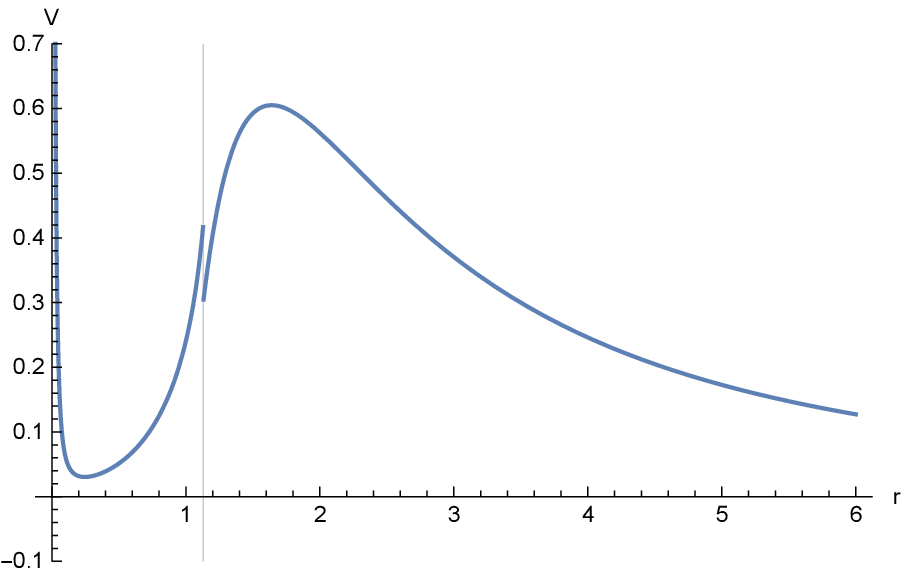}\includegraphics[width=0.8\columnwidth]{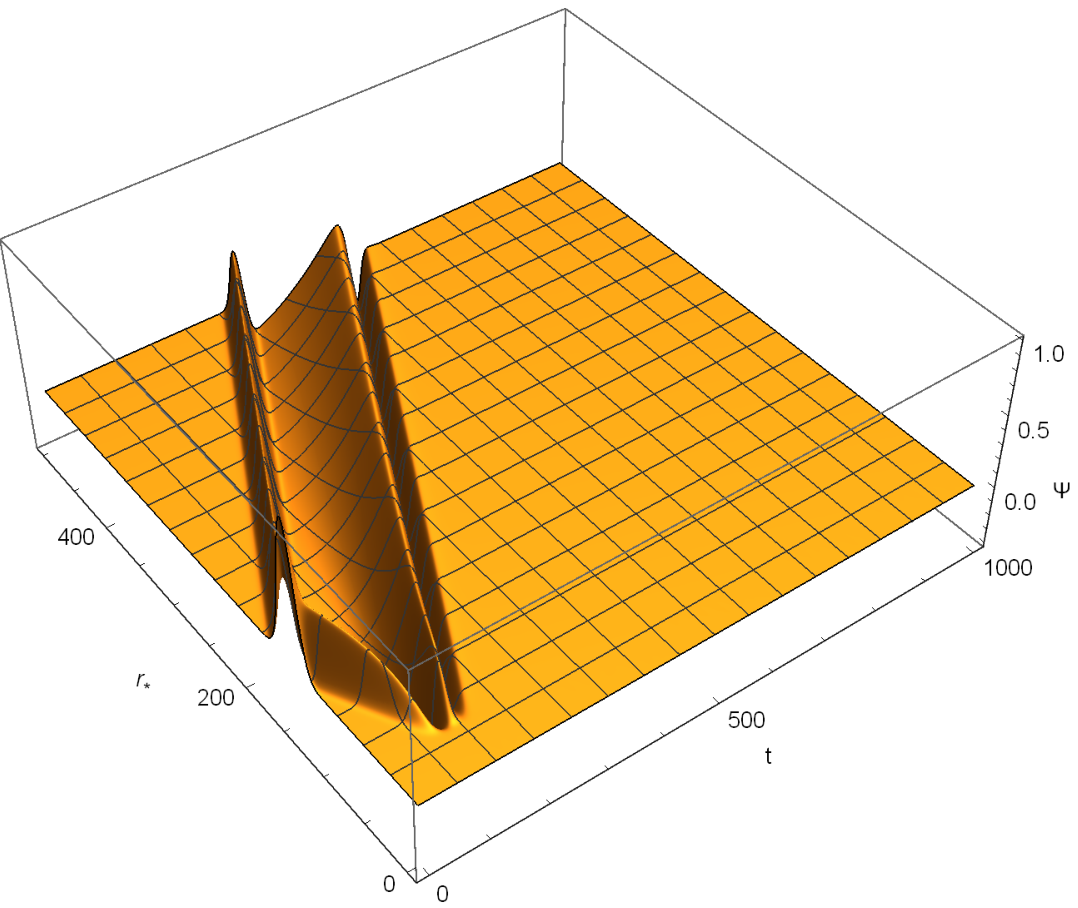}
\includegraphics[width=0.8\columnwidth]{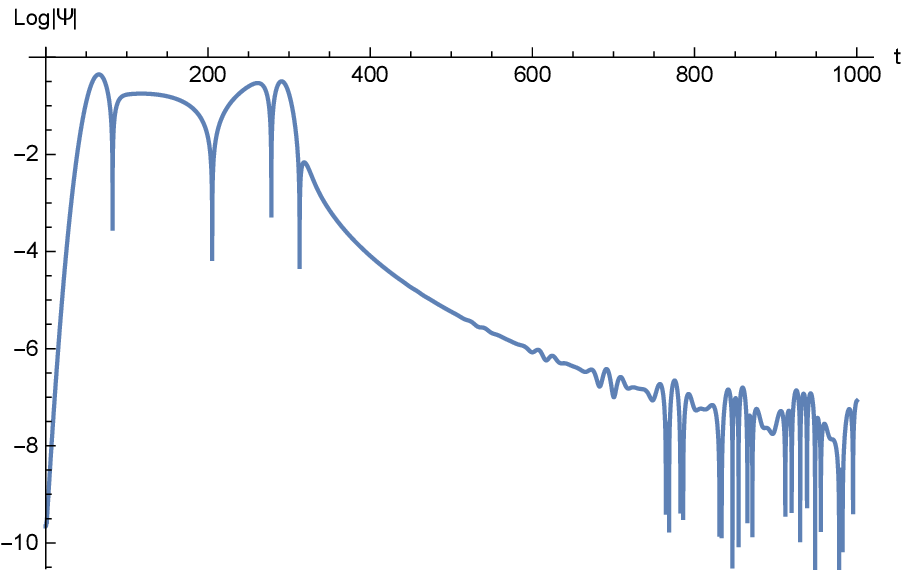}\includegraphics[width=0.8\columnwidth]{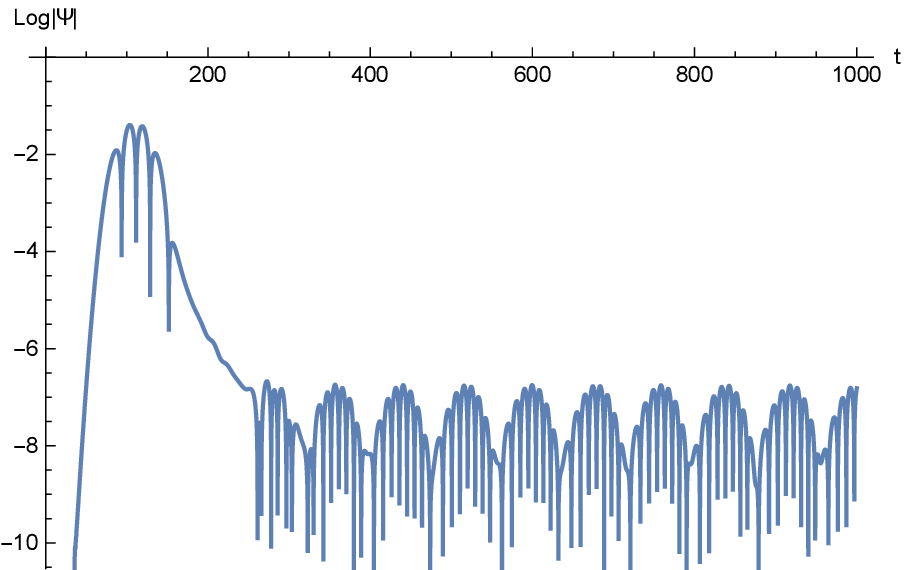}
\caption{The temporal evolution of the axial gravitational perturbations in the star with the surface $r_b=1.13$ ($r_{b*}=36.7$).
Top-left: The effective potential. The maximal value of potential is located at $r=1.64039$ ($r_*=38.8$), outside the star's surface.
Top-right: The two-dimensional profile of the temporal evolution, where the initial conditions read $\Psi(t=0)=e^{-(r_*-150)^2/200}$ and $\partial_{t}\Psi(t=0)=0$.
Due to the scale of the plot, the echoes are not visable in the two-dimensional plot.
Bottom-left: The temporal evolution evaluated at $r=173.937$ ($r_*=216.7$).
Bottem-right: The temporal evolution evaluated at $r=7.256 $ ($r_*=46.7$).}
\lb{Fig4}
\end{figure*}

In FIG.~4, we show the temporal evolutions of the axial gravitational perturbations in a star of uniform density as the simplest theoretical model for the neutron star.
The effective potential is featured by a valley between two maxima located at the center and the maximum of the potential.
As a result, the GWs are bounced back and forth in the valley, and subsequently, the echoes are produced in the late stage of dissipative oscillations.
Although the observer is located further away from the star, the echoes eventually leak out and appear in the signals.

\begin{figure*}[tbp]
\centering
\includegraphics[width=0.8\columnwidth]{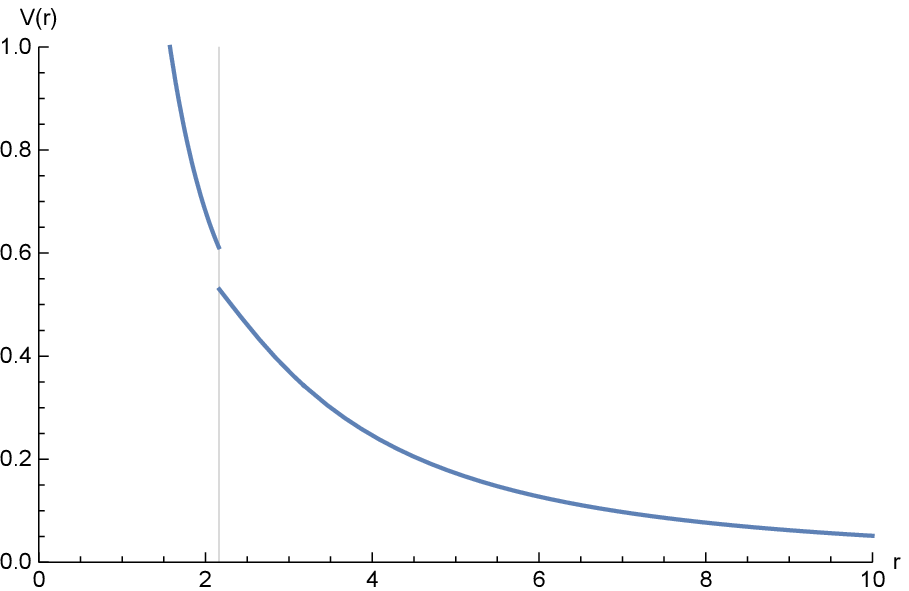}\includegraphics[width=0.8\columnwidth]{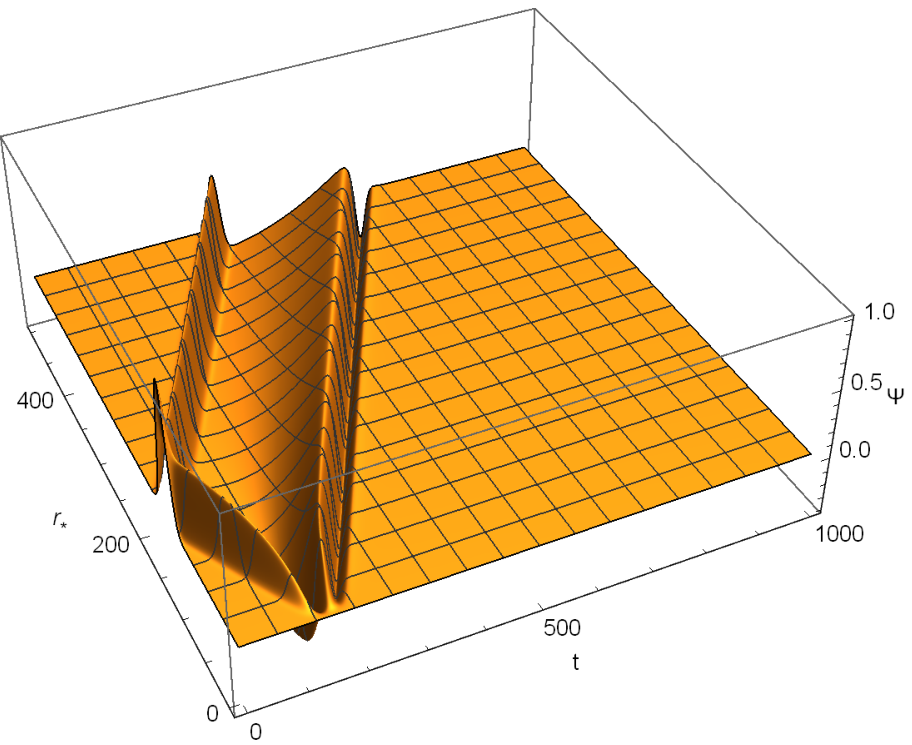}
\includegraphics[width=0.8\columnwidth]{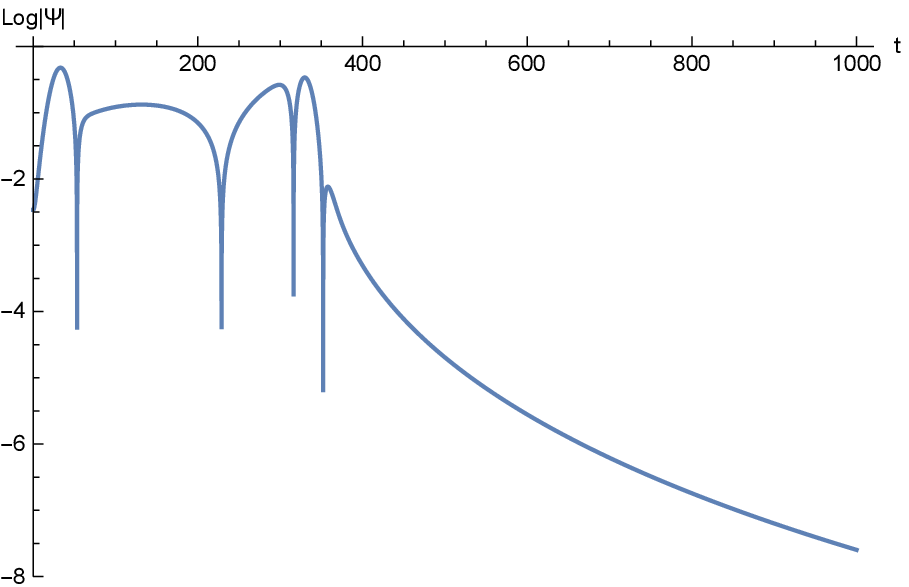}\includegraphics[width=0.8\columnwidth]{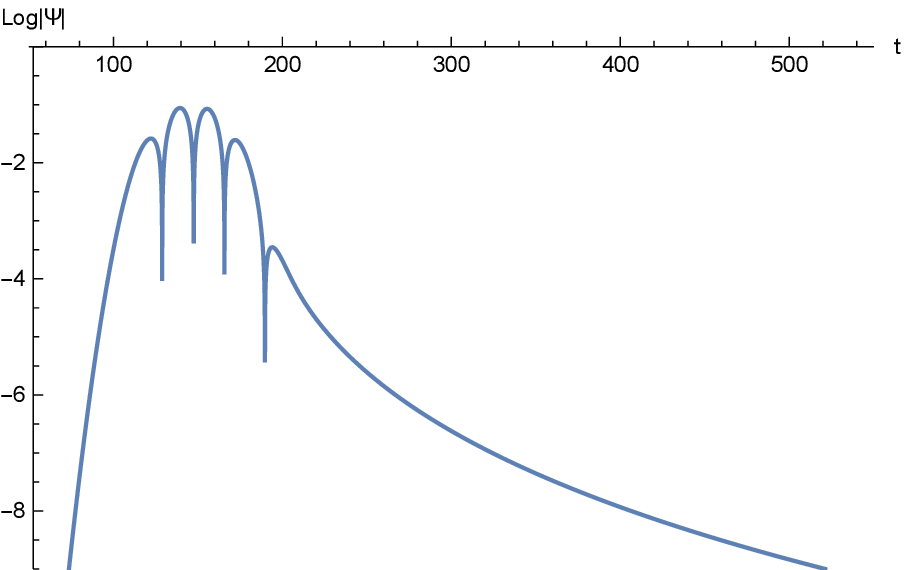}
\caption{The same as FIG.~4 but for $r_b=2.162$ ($r_{b*}=37$).
The bottom-left plot is evaluated at $r=177.141$ ($r_*=183.7$), the bottom-right one is evaluated at $r=10.104$ ($r_*=13.7$).}
\lb{Fig5}
\end{figure*}

As a comparison, as shown in FIG.~5, one similarly carries out the calculations by placing the star's surface outside the ``original'' maximum of the effective potential.
However, as shown in the top-left plot, the valley of the effective potential disappears in this case.
This is because the matter distribution inside the star dictates that the effective potential decreases monotonically with increasing radial coordinates.
Subsequently, one does not observe any echo in the resultant temporal evolution.

\section{Conclusion}

Using an FDM scheme with appropriate boundary conditions and treatment for the discontinuous boundary, we studied the temporal evolution of axial gravitational perturbations in black holes and uniform stars.
The results are shown to be consistent with causality and flow conservation.
In particular, echoes are observed in the late-stage of the star QNM oscillations, which is understood to be caused by the valley in the effective potential.
It is interesting to note that the echo period is numerically consistent with twice the distance between the center of the star and the maximum of the effective potential, while the discontinuity at the star's surface does not play a role~\cite{Hechoes}.
We plan to continue exploring the related topics in the future.

\section*{\bf Acknowledgements}
We gratefully acknowledge the financial support from the National Natural Science Foundation of China (NNSFC) under contract No. 42230207.
This research is also supported by from the Brazilian Agencies
Funda\c{c}\~ao de Amparo \`a Pesquisa do Estado de S\~ao Paulo (FAPESP),
Funda\c{c}\~ao de Amparo \`a Pesquisa do Estado do Rio de Janeiro (FAPERJ),
Conselho Nacional de Desenvolvimento Cient\'{\i}fico e Tecnol\'ogico (CNPq),
Coordena\c{c}\~ao de Aperfei\c{c}oamento de Pessoal de N\'ivel Superior (CAPES),
A part of this work was developed under the project Institutos Nacionais de Ciências e Tecnologia - Física Nuclear e Aplicações (INCT/FNA) Proc. No. 464898/2014-5.
We also acknowledge the support from the Center for Scientific Computing (NCC/GridUNESP) of the S\~ao Paulo State University (UNESP).

\onecolumngrid

\end{document}